\documentclass[letterpaper]{jpconf}
\usepackage{graphicx}
\begin{document}
\title{PNJL model for adjoint fermions}

\author{Hiromichi Nishimura and Michael C. Ogilvie}

\address{Department of Physics, Washington University, St.\ Louis, Missouri 63130, USA
}

\ead{hnishimura@physics.wustl.edu}

\begin{abstract}
Recent work on QCD-like theories has shown that the addition of adjoint
fermions obeying periodic boundary conditions to gauge theories on
$R^{3}\times S^{1}$ can lead to a restoration of center symmetry
and confinement for sufficiently small circumference $L$ of $S^{1}$.
At small $L$, perturbation theory may be used reliably to compute
the effective potential for the Polyakov loop $P$ in the compact direction.
Periodic adjoint fermions act in opposition to the gauge fields, which
by themselves would lead to a deconfined phase at small $L$. In order
for the fermionic effects to dominate gauge field effects in the effective
potential, the fermion mass must be sufficiently small.
This indicates
that chiral symmetry breaking effects are potentially important. We
develop a Polyakov-Nambu-Jona Lasinio (PNJL) model which combines
the known perturbative behavior of adjoint QCD models at small $L$
with chiral symmetry breaking effects to produce an effective potential
for the Polyakov loop $P$ and the chiral order parameter $\bar{\psi}\psi$.
A rich phase structure emerges from the effective potential. Our results 
\cite{Nishimura:2009me}
are consistent with the recent lattice simulations of Cossu and D'Elia \cite{Cossu:2009sq},
which found no evidence for a direct connection between 
the small-$L$ and large-$L$ confining regions.
Nevertheless, the
two confined regions are connected indirectly if an extended field
theory model with an irrelevant four-fermion interaction is considered. Thus
the small-$L$ and large-$L$ regions are part of a single confined phase.\end{abstract}

\section{Introduction}

Recent progress in the study of QCD-like gauge theories has revealed
that a confined phase can exist under certain conditions
when one or more spatial directions
are compactified and small \cite{Unsal:2007vu,Myers:2007vc}.
This is surprising, because a small compact direction in Euclidean
time gives rise to a deconfined phase for $SU(N)$ gauge theories at high temperatures \cite{Gross:1980br,Weiss:1980rj}.
It is also intriguing, because one or more small compact directions give
rise to a small effective coupling constant if
the theory is asymptotically free. 
Thus we now have four-dimensional field theories in which confinement
holds, and holds under circumstances where semiclassical methods may
be reliably applied. 
At present, there are two methods known for achieving this. The first method directly
modifies the gauge action with terms nonlocal in the compact direction(s)
\cite{Myers:2007vc},
while the second adds adjoint fermions with periodic boundary conditions
in the compact direction(s) \cite{Unsal:2007vu}, which is our subject here.

Confinement in $SU(N)$ gauge theories is associated with an
unbroken global center symmetry, which is $Z(N)$ for $SU(N)$.
The order parameter for $Z(N)$ breaking in the compact direction is the
Polyakov loop, $P\left(\vec{x}\right)=\mathcal{P}\exp\left[i\int_{0}^{L}dx_{4}A_{4}\left(x\right)\right]$, which is the path-ordered exponential of the gauge
field in the compact direction.
The trace of $P$ in a representation $R$ represents the insertion
of a heavy fermion in that representation into the system. 
Unbroken
$Z(N)$ symmetry implies $\langle Tr_{F}P\rangle=0$ in the confined
phase, and correspondingly $\left\langle Tr_{F}P\right\rangle \ne0$ holds
in the deconfined phase where $Z(N)$ symmetry is broken.

In the case of adjoint fermions with periodic boundary conditions on $R^{3}\times S^{1}$, $Z(N)$ symmetry is restored if the circumference $L$ of $S^{1}$ is sufficiently
small and the mass $m$ of the adjoint fermions is sufficiently light  \cite{Unsal:2007vu,Myers:2009df}. If $mL$ is sufficiently small,
the effective potential has a global minimum when the Polyakov loop
eigenvalues are uniformly spaced around the unit circle. This is the
unique $Z(N)$-symmetric solution for $P$. 
Experience
with phenomenological models 
\cite{Gocksch:1984yk,Fukushima:2003fw}
suggests that in fact it is the constituent
mass which is relevant in determining the size of the fermionic contribution
to the effective potential for $P$. 

In order to explore the interrelationship of confinement and chiral
symmetry breaking, we use a generalization of Nambu-Jona Lasinio models
known as Polyakov-Nambu-Jona Lasinio (PNJL) models  \cite{Fukushima:2003fw}. 
In NJL models,
a four-fermion interaction induces chiral symmetry breaking. There
has been a great deal of work on NJL models, both as phenomenological
models for hadrons and as effective theories of QCD 
\cite{Klevansky:1992qe,Hatsuda:1994pi}.
NJL models have been used to study hadronic physics at
finite temperature, but they include only chiral symmetry restoration,
and do not model deconfinement. This omission is rectified by the
PNJL models, which include both chiral restoration and deconfinement.
The earliest model of this type was derived from strong-coupling lattice
gauge theory \cite{Gocksch:1984yk}, but later work on continuum models have
proven to be extremely powerful in describing the finite-temperature
QCD phase transition \cite{Fukushima:2003fw}. In PNJL models, fermions with
NJL couplings move in a nontrivial Polyakov loop background, and
the effects of gluons at finite temperature is modeled in a semiphenomenological
way. We will develop a model of this type for both fundamental
and adjoint fermions below.

Recent lattice simulations by Cossu and D'Elia \cite{Cossu:2009sq}
have confirmed the existence of the small-$L$ confined region in
$SU(3)$ lattice gauge theory with two flavors of adjoint fermions,
and we will focus on this case in our analysis.
Even if
the small-$L$ confined region exists and is accessible in lattice
simulations, it is not necessarily the same phase as found for large
$L$. Put slightly differently, we would like to know if the small-$L$
and large-$L$ confined regions are smoothly connected, and thus represent
the same phase. Our main result
will be a phase diagram for adjoint periodic QCD for all values of $L$, 
obtained using a PNJL model.
On the way to this goal, we will use as tests of our model
both standard QCD with fundamental fermions 
and adjoint QCD with the usual antiperiodic boundary
conditions for fermions. Our principal
tool will be the effective potential for the chiral symmetry order
parameter $\bar{\psi}\psi$ and the deconfinement order parameter
$P$. For a more detailed discussion, see reference \cite{Nishimura:2009me}.

\section{Effective potential}

\subsection{Fermionic contribution}

We take the fermionic part of the Lagrangian of our PNJL model to be 
\cite{Fukushima:2003fw,Klevansky:1992qe,Hatsuda:1994pi}
\begin{equation}
L_{F}=\bar{\psi}\left(i\gamma\cdot D-m_{0}\right)\psi+\frac{g_{S}}{2}\left[\left(\bar{\psi}\lambda^{a}\psi\right)^{2}+\left(\bar{\psi}i\gamma_{5}\lambda^{a}\psi\right)^{2}\right]+g_{D}\left[\det\bar{\psi}\left(1-\gamma_{5}\right)\psi+h.c.\right]\end{equation}
where $\psi$ is associated with $N_{f}$ flavors of Dirac fermions
in the fundamental or adjoint representation of the gauge group $SU(N)$.
The $\lambda^{a}$'s are the generators of the flavor symmetry group
$U(N_{f})$ and $ \lambda_{ij}^{a}\lambda_{kl}^{a}=2\delta_{il}\delta_{jk}$; $g_{S}$ represents the strength of the four-fermion
scalar-pseudoscalar coupling and $g_{D}$ fixes the strength of an
anomaly induced term. For simplicity, we take the Lagrangian mass
matrix $m_{0}$ to be diagonal: $\left(m_{0}\right)_{jk}=m_{0j}\delta_{jk}$.
The covariant derivative $D_{\mu}$ couples the fermions to a background Polyakov
loop via the component of the gauge field in the compact direction.

It is generally convenient to use the language of finite temperature to describe
both the case of  finite temperature, $\beta^{-1}=T>0$, with antiperiodic boundary conditions,
and the case of a periodic spatial direction, $L<\infty$.
The zero-temperature contribution to the fermionic effective potential is given by
\begin{equation}
V_{F0}\left(m,m_{0}\right)=\sum_{j}g_{S}\sigma_{j}^{2}+2g_{D}\left(N_{f}-1\right)\prod_{j}\sigma_{j}
-2d_{R}\sum_{j=1}^{N_{f}}\int^{\Lambda}\frac{d^{3}k}{(2\pi)^{3}}\omega_{k}^{\left(j\right)}
\end{equation}
where $\omega_{k}^{\left(j\right)}=\sqrt{k^{2}+m_{j}^{2}}$, $\sigma_{j}=\langle \bar{\psi}_j\psi_j \rangle$, $m_{j}=m_{0j}-2g_{S}\sigma_{j}-2g_{D}\prod_{k\ne j}\sigma_{k}$ is a constituent mass, and the
constant $d_{R}$ is the dimensionality of the color representation,
$N$ for the fundamental and $N^{2}-1$ for the adjoint. The last term,
representing a sum of one-loop diagrams, is regularized by three-dimensional momentum space cutoff, $\Lambda$  \cite{Klevansky:1992qe}.

In PNJL models, the finite-temperature contribution from the fermion
determinant depends on the background Polyakov loop. It is convenient
to work in a gauge where the temporal component of the background
gauge field, $A_{4}(\vec{x},\, t)$, is constant and diagonal. The
covariant derivative then becomes $\gamma\cdot D=\gamma\cdot\partial-i\gamma^{4}A_{4}$.
The one-loop free energy of fermions in a representation $R$ of $SU(N)$
gauge theory with zero chemical potential can be expanded in terms of modified Bessel functions
\begin{equation}
V_{FL}\left(P,m\right)=\sum_{j}\frac{2m_{j}^{2}}{\pi^{2}L^{2}}\sum_{n=1}^{\infty}\frac{(\pm 1)^{n}Tr_{R}P^{n}}{n^{2}}K_{2}\left(nL m_{j}\right)\end{equation}
which is rapidly convergent for all values of the mass
\cite{Meisinger:2001fi}. The plus sign is used for periodic boundary conditions and minus for antiperiodic.
In what follows, we will take $N_{f}=2$, and take the masses $m_{0j}$
to be equal to a common mass which we also write as $m_{0}$. In this
case, the contribution to the effective action from $g_{S}$ and $g_{D}$ has
the same form. It is convenient to take $g_{D}=0$, and also to write
the common constituent mass as $m=m_{0}-2g_{S}\sigma$ 
\cite{Klevansky:1992qe,Hatsuda:1994pi}.
There is a possibility of directly
modifying the strength of chiral symmetry breaking by adding additional
couplings compatible with all symmetries have been added. In the case of adjoint fermions
with periodic boundary conditions, the ability to freely vary $g_S$ 
allows a clear connection between the large-$L$ and small-$L$
confining regions of the phase diagram.

\subsection{Gluonic contribution}

The boundary conditions for the gauge bosons
are periodic in all cases considered here, so $L$ and $\beta$
may be used equivalently in the gluonic sector.
The one-loop finite-temperature free energy in a background Polyakov loop is given by an expression
similar to the one for fermions
\begin{equation}
V_{g-1\,loop}\left(P\right)=2\, Tr_{A}[\frac{1}{L}\int\frac{d^{3}k}{(2\pi)^{3}}ln(1-Pe^{-L\Omega_{k}})]\end{equation}
where we have inserted a mass parameter in $\Omega_{k}=\sqrt{k^{2}+M^{2}}$
for purely phenomenological reasons explained below.

The Polyakov loop in the fundamental representation
of $SU(3)$ can be diagonalized by a gauge transformation
and written as $P_{jk}=\exp (i\phi_{j})\delta_{jk}$
with two independent angles. With the use of $Z(3)$ symmetry,
it is sufficient to consider the case where
$\left\langle Tr_{F}P\right\rangle $ is real.
Thus we consider only
diagonal, special-unitary matrices with real trace, which may be parametrized
by taking $\phi_{1}=\phi$, $\phi_{2}=-\phi$, and $\phi_{3}=0$,
or $P=diag\left[e^{i\phi},e^{-i\phi},1\right]$ with $0\le\phi\le\pi$.
The unique set of $Z(3)$-invariant eigenvalues are obtained for $\phi=2\pi/3$.
For $SU(3)$, we can write the gluonic effective potential in a high temperature expansion in terms of $\phi$
 \cite{Meisinger:2001cq}
\begin{equation}
V_{g}\left(P\right)=\left(\frac{3\phi^{2}}{2\pi^{2}}-\frac{2\phi}{\pi}+\frac{2}{3}\right)\frac{M^{2}}{L^{2}}+\frac{1}{L^{4}}\left(\frac{135\phi^{4}-300\pi\phi^{3}+180\pi^{2}\phi^{2}-16\pi^{4}}{90\pi^{2}}\right).\end{equation}
We will set the mass scale $M$ by requiring that $V_{g}$ yields the correct
deconfinement temperature for the pure gauge theory,
with a value of  $T_d\approx270\, MeV$. This gives $M=596\, MeV$ \cite{Meisinger:2001cq}.
We stress that the mass parameter $M$ should not be interpreted
as a gauge boson mass, nor do we limit ourselves to $ML\ll1$. The
crucial feature of this potential is that for sufficiently large values
of the dimensionless parameter $ML$, the potential leads to a $Z(N)$-symmetric,
confining minimum for $P$ \cite{Myers:2009df,Meisinger:2009ne}. 
On the other hand, for small values
of $ML$, the pure gauge theory will be in the deconfined phase. It
will be important later that $V_g$ is a good
representation of the gauge boson contribution for high temperatures;
in other PNJL models, the gauge boson contribution has sometimes
been chosen so as to be valid over a more narrow range of temperatures.

\section{Fundamental Fermions}

\begin{figure}[h]
\begin{minipage}{3in}
\includegraphics[width=3in]{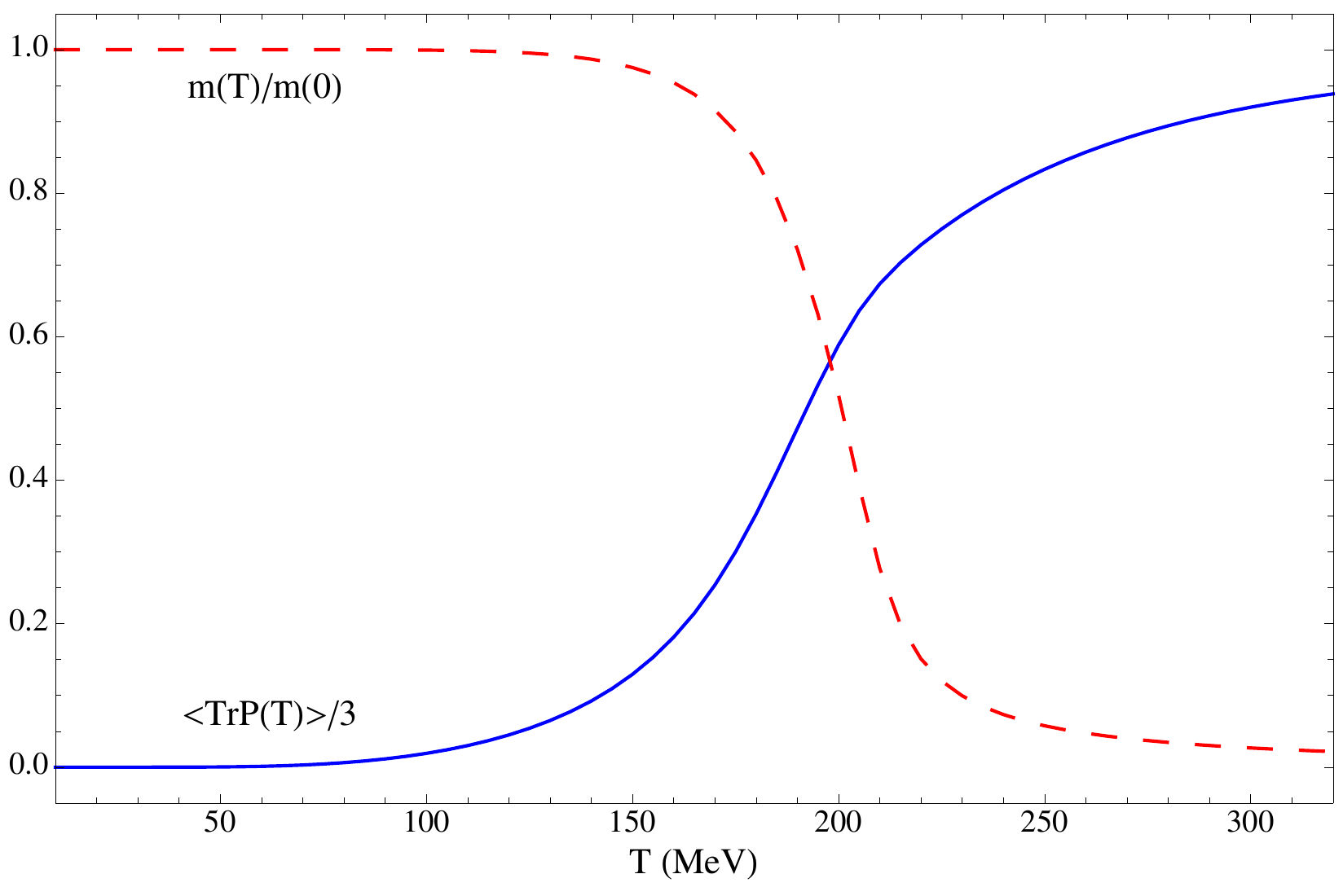}
\caption{The constituent mass $m$ and $\langle Tr_F P \rangle$ for two-flavor
QCD with fundamental representation fermions with antiperiodic boundary conditions
as a function of temperature.}
\label{fig:TvsOPs_FundABC}
\end{minipage}\hspace{2pc}%
\begin{minipage}{3in}
\includegraphics[width=3in]{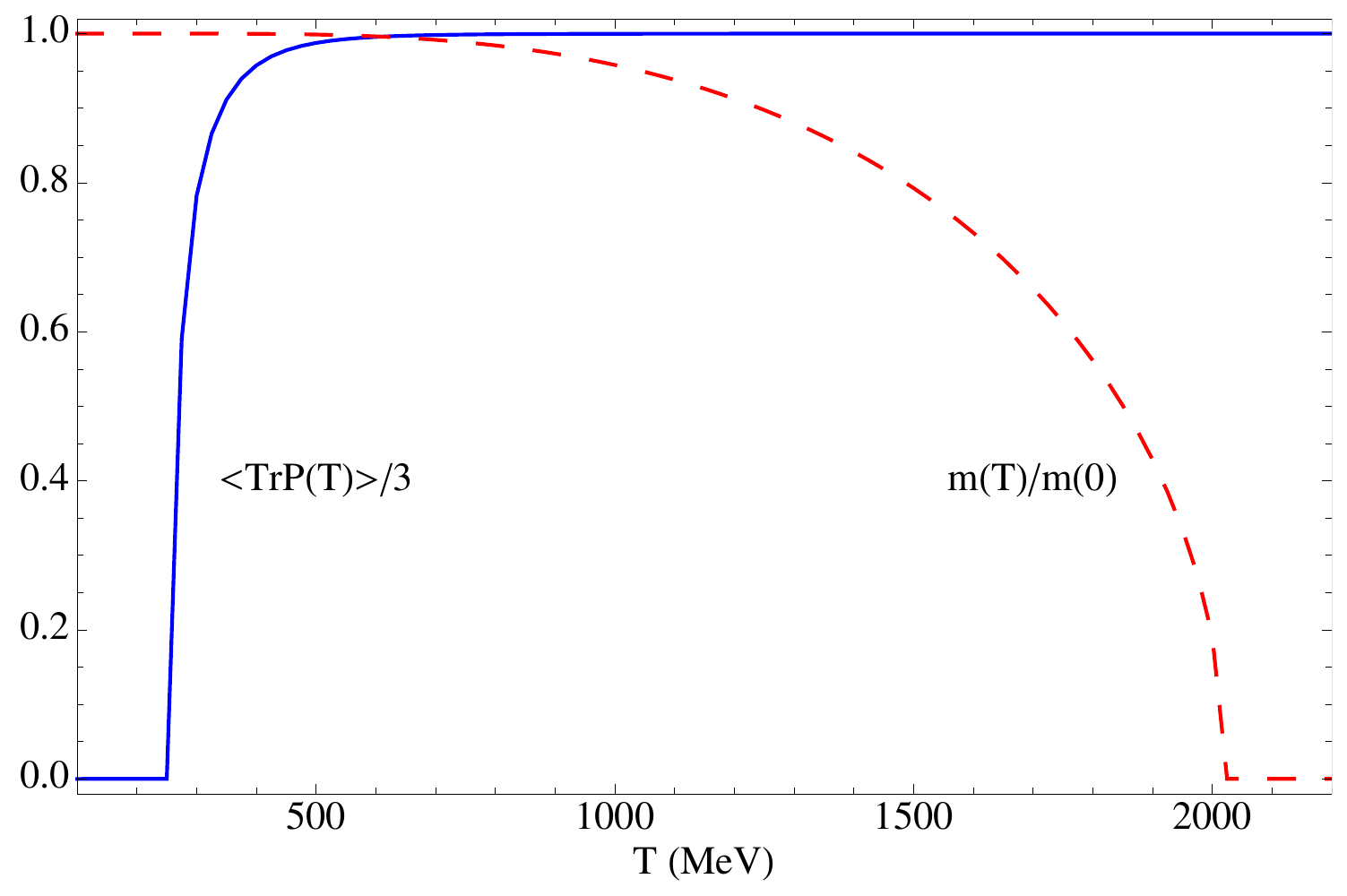}
\caption{The constituent mass $m$ and $\langle Tr_F P \rangle$ for two-flavor
QCD with adjoint representation fermions with antiperiodic boundary conditions
as a function of temperature.}
\label{fig:TvsOPs_AdjABC}
\end{minipage} 
\end{figure}

As a test of all the components of the effective potential we have
assembled, we consider the case of two flavors of fundamental fermions
at finite temperature. A very common choice of zero-temperature
parameters for two degenerate light flavors is $m_{0}=5.5MeV,$ $\Lambda=631.4\,MeV,$
and $g_{S}=2\times5.496GeV^{-2}$
\cite{Fukushima:2003fw,Hatsuda:1994pi}.
We will use these parameters, augmented by
the gluonic sector parameter $M=596\, MeV$ discussed in the
previous section.
In Figure \ref{fig:TvsOPs_FundABC}, we show the constituent mass $m$ and 
Polyakov expectation value $\langle Tr_F P \rangle$ 
as a function of temperature,
normalized by dividing by their values at $T=0$ and $T=\infty$, respectively.
The behavior in the crossover region is very similar
to the results of Fukushima \cite{Fukushima:2003fw}, and
shows the explanatory power of PNJL models. The constituent mass $m$
is heavy at low temperatures, due to chiral symmetry breaking. The
larger the constituent mass, the smaller the $Z(3)$ breaking effect
of the fermions. On the other hand, a small value for $\left\langle Tr_{F}P\right\rangle $
reduces the effectiveness of finite-temperature effects in restoring chiral symmetry.
These synergistic effects combine in the case of fundamental representation
fermions to give a single crossover temperature at which both order
parameters are changing rapidly, in agreement with lattice simulations.

\section{Adjoint Fermions with antiperiodic boundary conditions}

Adjoint $SU(3)$ fermions at finite temperature show a completely
different behavior in lattice simulations from fundamental fermions. 
Because the adjoint fermions
respect the $Z(3)$ center symmetry, there is a true deconfinement
transition where $Z(3)$ spontaneously breaks.
Lattice simulations have shown that chiral symmetry is
restored at a substantially higher temperature than
the deconfinement temperature \cite{Karsch:1998qj,Engels:2005te}.
The $T=0$ parameters
needed are $g_{S}$ and $\Lambda$. 
Rather than work directly with
$g_{S}$, we will consider the dimensionless coupling $\kappa=g_{S}\Lambda^{2}$.
A given ratio of $m(T=0)/\Lambda$ determines
the value of $\kappa$, and vice versa.
The value of $\Lambda$ is determined by the requirement that
$T_{c}/T_{d}$ is near $7.8$ \cite{Engels:2005te}. This in turn determines the value of the constituent
mass for all $T$.

The ratio $m(T=0)/\Lambda$
should be less than one in
order for the cutoff theory to be meaningful.
In the case of fundamental fermions, this ratio
is relatively large, on the order of $0.5$.
We have generally found that for adjoint fermions
a larger ratio of $m(T=0)/\Lambda$
with $T_{c}/T_{d}$ fixed
implies a larger value of $m(T=0)$.
We will work with the representative
case of $m_{0}=0$ 
and $m(T=0)/\Lambda=0.1$. 
This gives $\Lambda=23.22\,GeV$ and thus 
$m(T=0)=2.322\,GeV$, with $\kappa=1.2653$. 
For comparison, the critical value of $\kappa$, $\kappa_c$,
below which $m(T=0)=0$, is $\pi^2/8\simeq 1.234$.
In Figure \ref{fig:TvsOPs_AdjABC},
we show the constituent mass $m$ and 
Polyakov expectation value $\langle Tr_F P \rangle$ 
as a function of temperature,
normalized by dividing by their values at $T=0$ and $T=\infty$, respectively.
We see that the deconfinement temperature $T_d$
is very close to its value in the pure gauge theory,
due to the large adjoint fermion constituent mass.
The transition is first order.
The constituent mass $m$ has a slow decline
to a second-order transition at a substantially
higher temperature, as indicated by lattice simulations 
\cite{Karsch:1998qj,Engels:2005te}.

\section{Adjoint Fermions with periodic boundary conditions}

\begin{figure}[h]
\begin{minipage}{3in}
\includegraphics[width=3in]{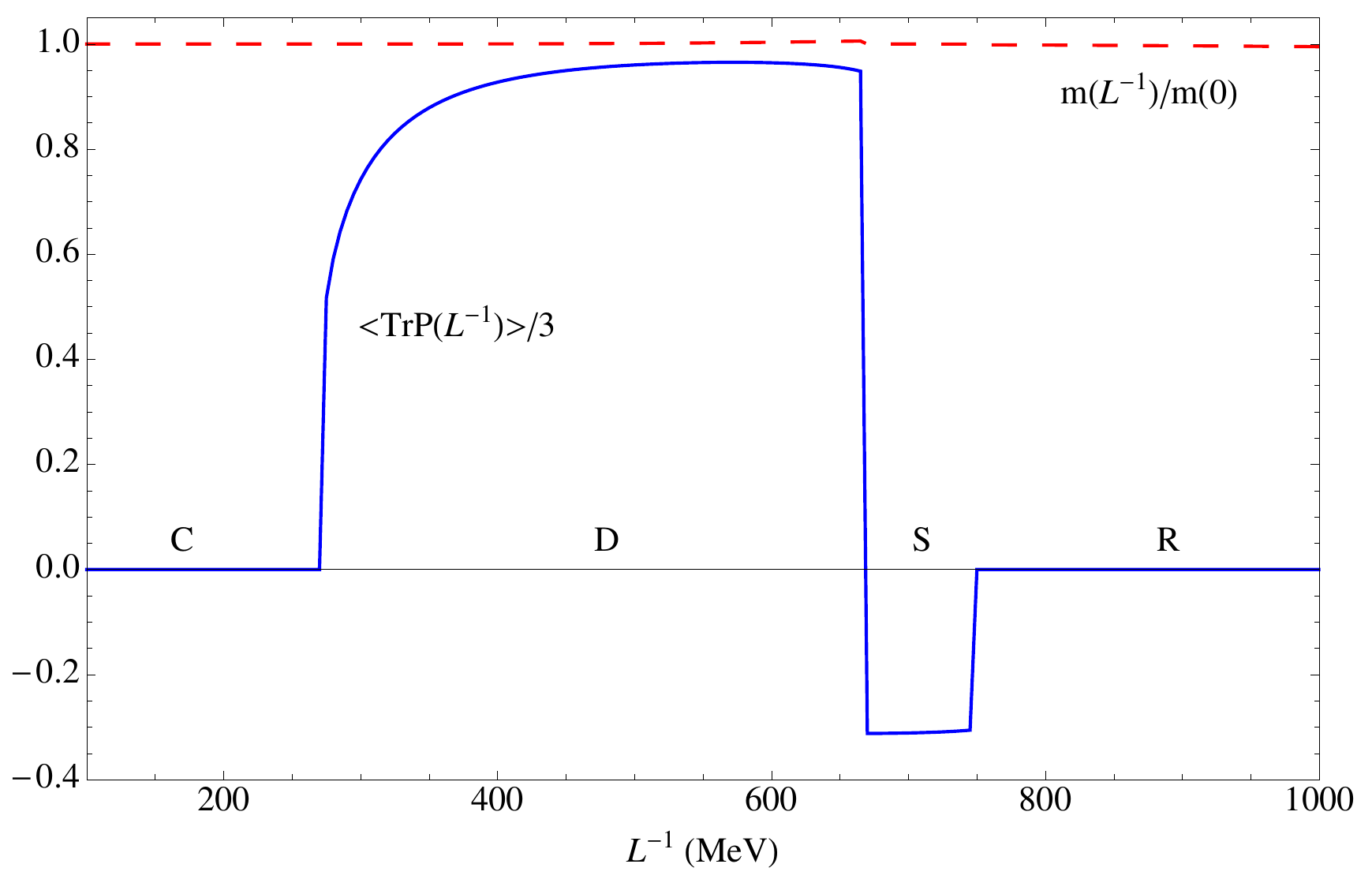}
\caption{The constituent mass $m$ and $\langle Tr_F P \rangle$ for two-flavor
QCD with adjoint representation fermions with periodic boundary conditions
as a function of $L^{-1}$. C, D, S and R refer to the confined, deconfined, skewed,
and reconfined phase, respectively.}
\label{fig:TvsOPs_AdjPBC}
\end{minipage}\hspace{2pc}%
\begin{minipage}{3in}
\includegraphics[width=3in]{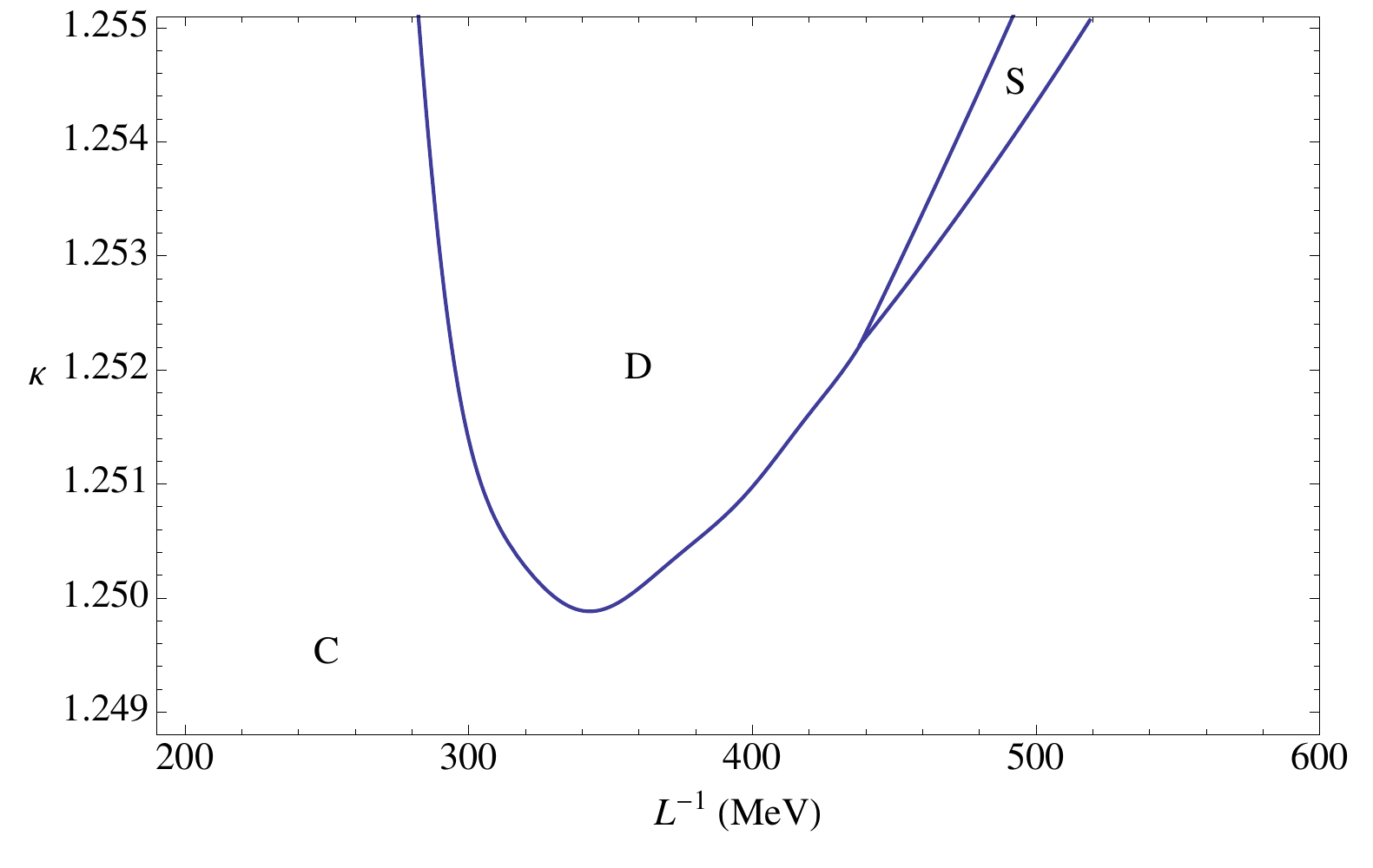}
\caption{The phase diagram for two-flavor
QCD with adjoint representation fermions with periodic boundary conditions
in the $L^{-1}$-$\kappa$ plane. $\kappa=g_{S}\Lambda^2$. C, D, and S refer to the confined, deconfined, and skewed phase, respectively.}
\label{fig:PhaseDiagram_Fit}
\end{minipage} 
\end{figure}

We consider the behavior of $m$ and $Tr_{F}P$ with periodic
fermions using the same parameters we used for the antiperiodic case.
Figure \ref{fig:TvsOPs_AdjPBC} shows the behavior of $m$ 
and $\langle Tr_F P\rangle$ as a function of $L^{-1}$ for
the $m(L=\infty)/\Lambda=0.1$ parameter set,
with $m_0=0$.
We see that chiral symmetry breaking
persists at $L^{-1}=10\, GeV$, which is much higher than the
chiral restoration temperature for antiperiodic fermions. 
The constituent mass
$m$ does fall eventually as $L^{-1}$ increases, and
chiral symmetry is ultimately restored,
but at a temperature on the order of $\Lambda$.
In Figure
\ref{fig:TvsOPs_AdjPBC}, $Tr_{F}P$ shows
three distinct phase transitions as
a function of $L^{-1}$. As $L^{-1}$ increases, the confined phase
gives way to the deconfined phase in a first-order phase transition.
Because the constituent mass of the fermions is large, the critical
value of $L^{-1}$ for this transition is approximately equal to $T_d$.
As $L^{-1}$ increases, there are two more first-order
transitions, from the deconfined phase to the skewed phase, and then
from the skewed phase to a small-$L$ confined phase we describe 
as reconfined.
The ordering of the phases seen in the behavior of $Tr_{F}P$ for
$m_{0}=0$ persists as $m_{0}$ is increased \cite{Nishimura:2009me}.

In Figure \ref{fig:PhaseDiagram_Fit}, 
we show the phase diagram in the $L^{-1}-\kappa$ plane,
obtained by numerically
minimizing $V_{eff}$. For most values of $\kappa$ larger than $\kappa_{c}$,
the confined large-$L$ phase and the reconfined phase at 
small $L$ are separated by three phase transitions as in Figure \ref{fig:TvsOPs_AdjPBC}. All of these
transitions are characterized by abrupt changes in $Tr_{F}P$, while
the chiral order parameter shows only a slow decrease with increasing
temperature. However, there is a narrow range of $\kappa$ between
approximately $1.250$ and $\kappa_{c}\simeq 1.234$ where confinement holds at all
temperatures, and chiral symmetry remains broken. In this extended
phase diagram, the confined and reconfined regions are smoothly connected. Although this connection appears only for small range of values, the corresponding range of constituent mass values is not necessarily small \cite{Nishimura:2009me}. 
Our results bear directly on the recent work by Cossu and D'Elia 
\cite{Cossu:2009sq}, in which they performed lattice simulations of two-flavor
$SU(3)$ gauge theory with periodic adjoint fermions. 

\section{Conclusions}

We have extended the PNJL treatment of $SU(3)$ gauge theories to
the case of two adjoint fermions with periodic boundary conditions on
$R^{3}\times S^{1}$. 
Our simple model reproduces the
known successes of PNJL models for fundamental fermions
while at the same time reproducing
the expected behavior at high temperatures
needed with adjoint fermions. 
The large separation between the deconfinement
transition and the chiral symmetry restoration transition for adjoint
fermion theories with antiperiodic boundary conditions requires a PNJL
model which reproduces the behavior of the pure gauge theory to much
smaller values of $L$ than have been considered before. 

The results for our $SU(3)$ PNJL model with two flavors of periodic
adjoint Dirac fermions can be summarized in the phase diagram in Figure \ref{fig:PhaseDiagram_Fit}.
They are completely compatible with the lattice
simulations of Cossu and D'Elia 
\cite{Cossu:2009sq}. If $m_{0}$ is
set to zero, there is a small region in the $L^{-1}-\kappa$ plane, lying
above $\kappa_{c}$, that connects the large-$L$ and small-$L$ confined
regions. Because the largest contribution to the constituent mass
$m$ is from chiral symmetry breaking, this behavior will persist
for some small range of nonzero $m_{0}$. Thus there is a single
confining region, accessible in principle in lattice simulations.

\end{document}